\documentclass[numberedappendix]{emulateapj}

\begin{document}

% document history:

\bibliographystyle{apj}

\shorttitle{Millimeter Moth}

\shortauthors{Ricarte et al.}

\slugcomment{Accepted for Publication in ApJ: July 11, 2013}

\title{
Resolving The Moth at Millimeter Wavelengths
}

\author{Angelo Ricarte\altaffilmark{1},
Noel Moldvai\altaffilmark{1},
A. Meredith Hughes\altaffilmark{2},
Gaspard Duch\^ene\altaffilmark{1,3},
Jonathan P. Williams\altaffilmark{4},
Sean M. Andrews\altaffilmark{5},
David J. Wilner\altaffilmark{5}
}
\altaffiltext{1}{University of California Berkeley, Department of Astronomy, 601 Campbell Hall, Berkeley, CA 94720}
\altaffiltext{2}{Wesleyan University Department of Astronomy, Van Vleck Observatory, 96 Foss Hill Drive, Middletown, CT 06459}
\altaffiltext{3}{UJF-Grenoble 1/ CNRS-INSU, Institut de Plan\'etologie et
d’Astrophysique de Grenoble (IPAG) UMR 5274, BP 53, 38041 Grenoble Cedex 9, France}
\altaffiltext{4}{Institute for Astronomy, University of Hawaii, etc.}
\altaffiltext{5}{Harvard-Smithsonian Center for Astrophysics, 60 Garden Street, Cambridge, MA 02138}

\begin{abstract}

HD 61005, also known as ``The Moth,'' is one of only a handful of debris disks
that exhibit swept-back ``wings'' thought to be caused by interaction with the
ambient interstellar medium (ISM).  We present 1.3\,mm Submillimeter Array (SMA)
observations of the debris disk around HD~61005 at a spatial resolution of 
1\farcs9 that resolve the emission from large grains for the first time.  The 
disk exhibits a double-peaked morphology at millimeter wavelengths, consistent 
with an optically thin ring viewed close to edge-on.  To investigate the disk 
structure and the properties of the dust grains we simultaneously model the 
spatially resolved 1.3\,mm visibilities and the unresolved spectral energy 
distribution.  The temperatures indicated by the SED are consistent with expected 
temperatures for grains close to the blowout size located at radii commensurate 
with the millimeter and scattered light data.  We also perform a visibility-domain 
analysis of the spatial distribution of millimeter-wavelength flux, incorporating 
constraints on the disk geometry from scattered light imaging, and find suggestive evidence of 
wavelength-dependent structure.  The millimeter-wavelength emission apparently 
originates predominantly from the thin ring component rather than tracing 
the ``wings'' observed in scattered light.  The implied segregation of large 
dust grains in the ring is consistent with an ISM-driven origin for the 
scattered light wings. 

\end{abstract}

\section{Introduction}

%Image
\begin{figure*}[ht]
\hspace{-10pt}
\begin{minipage}[c]{0.3\linewidth}
\centering 
\includegraphics[width=1.1\textwidth]{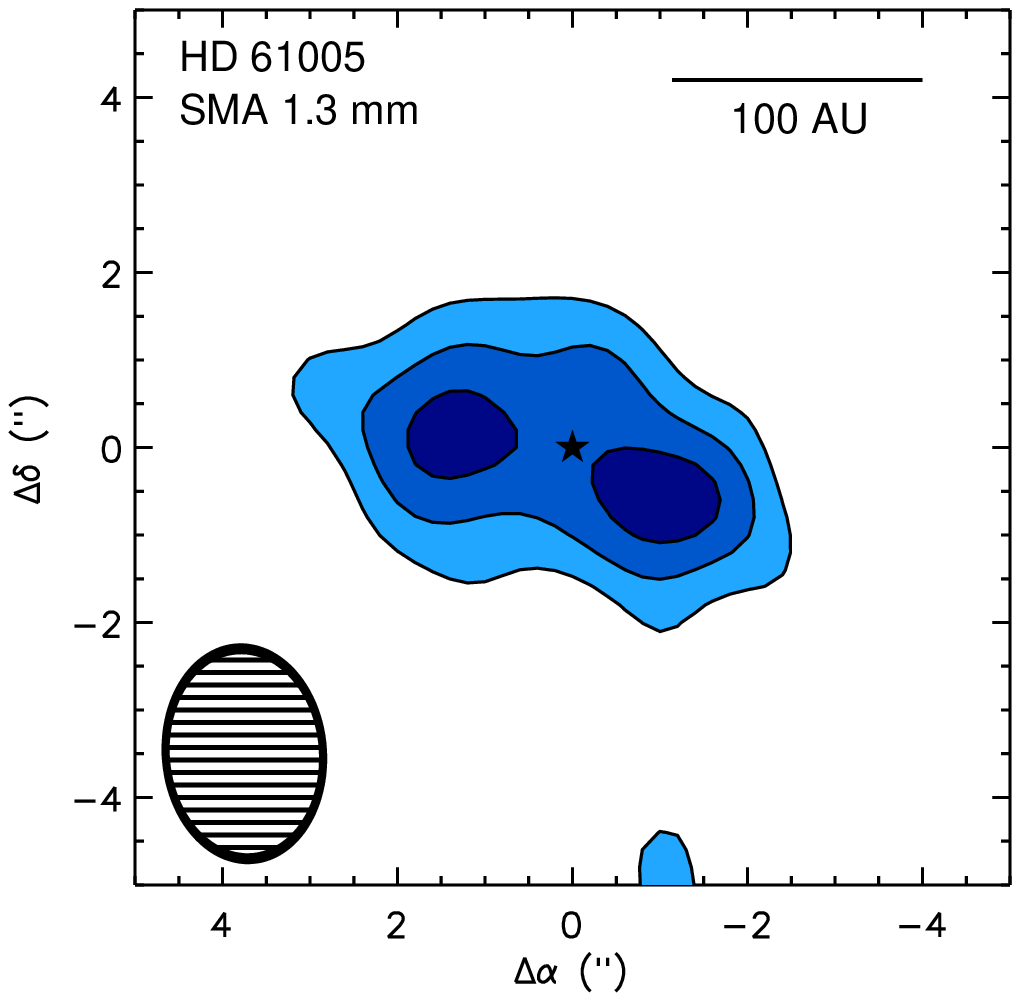} 
\end{minipage}
\hspace{-30pt}
\begin{minipage}[c]{0.3\linewidth}
\centering
\includegraphics[width=1.35\textwidth]{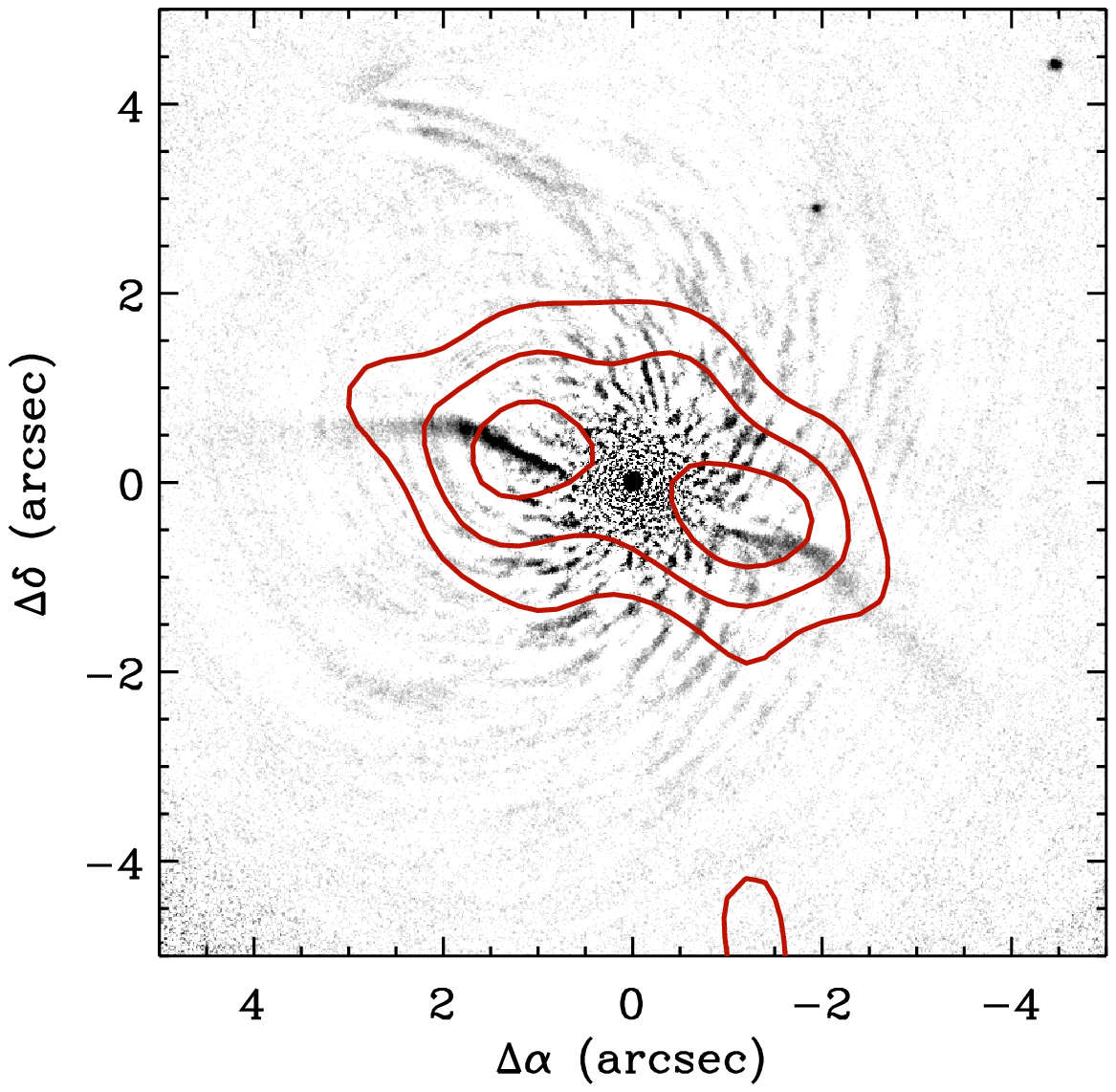}
\end{minipage}
\hspace{20pt}
\begin{minipage}[c]{0.3\linewidth}
\centering
\includegraphics[width=1.35\textwidth]{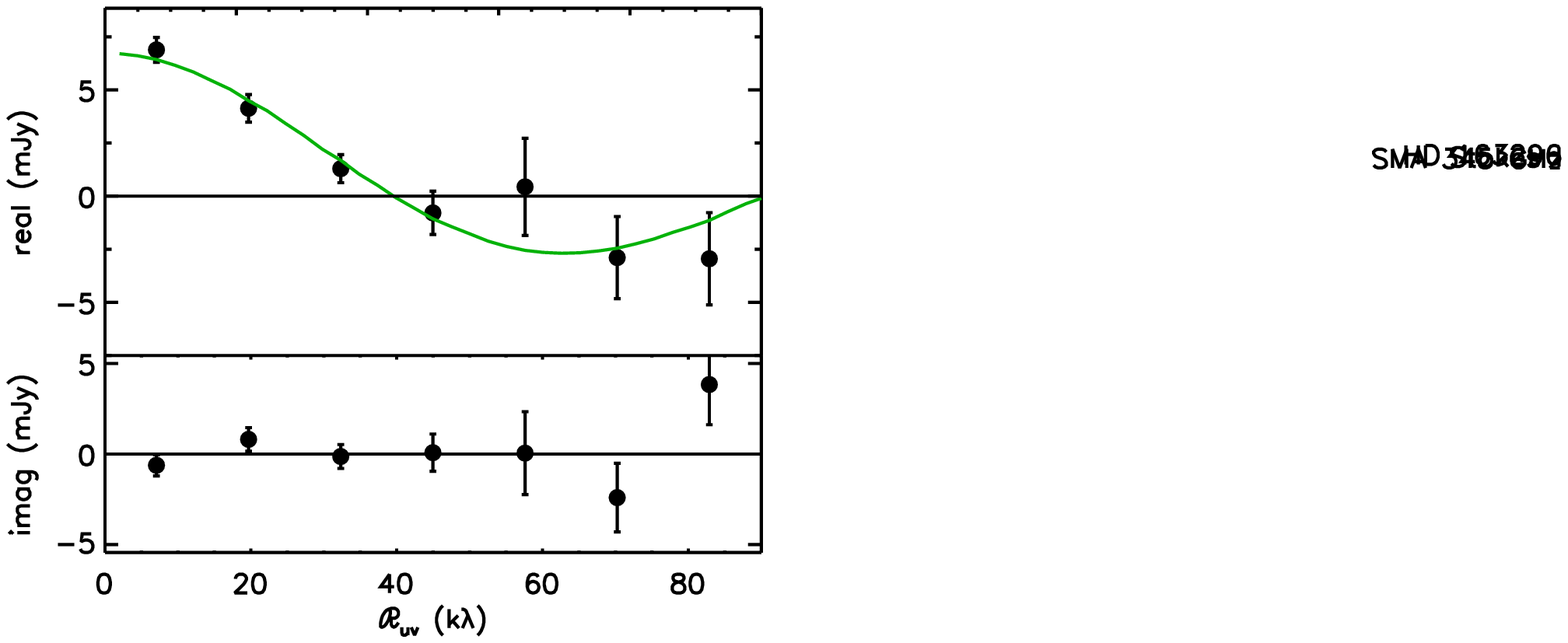}
\end{minipage}
\caption{Left:  SMA map of the 1.3\,mm emission from the debris disk around 
HD~61005.  The image has been reconstructed using all data with both array 
configurations.  The axes have been set such that the (0,0) position 
corresponds to the expected location of the star at the epoch of observation (with each
data set individually corrected for proper motion), and its position is marked 
by a star symbol.  Contours are drawn at [3,5,7]$\times$0.34\,mJy (the rms 
noise).  An ellipse indicating the size of the synthesized beam 
(2\farcs5$\times$1\farcs9 at a position angle of 4$^\circ$) is displayed in the 
bottom left corner.  The emission is resolved into two distinct peaks, 
consistent with the expected morphology for an optically thin debris disk 
viewed close to edge-on.  Center: 1.3\,mm contours overlaid onto H-band image 
\citet{bue10}.  These images were aligned by eye due to uncertainties in the 
pointing coordinates of the optical image.  The relatively low resolution of
the SMA data collects flux from along the disk major axis and causes the peaks
to appear closer together than the ansae of the scattered light ring.  Right: 1.3\,mm 
visibilities plotted as a function of the baseline length, deprojected to account for
the 84.3$^\circ$ inclination angle at which the source is viewed \citep[see, e.g.,][for 
a mathematical description of the abscissa]{hug08}.  The green line shows the visibilities
for the best-fit model.
  \label{plot:Image}}
\end{figure*}

Debris disks around main sequence stars provide an indicator that 
planet formation has proceeded at least to the scale of planetesimals.  
All three directly imaged planetary systems to date 
\citep[Fomalhaut, HR 8799, and $\beta$ Pictoris;][]{kal08,mar08,lag10} also 
host debris disks, and in two cases disk structure (eccentricity and warping) 
led to the prediction of at least one perturbing body in the system 
\citep[e.g.][]{hea00,bur95}.  Investigations of debris disk structure are 
therefore an important facet in our understanding of extrasolar planetary 
systems.  Millimeter-wavelength imaging plays an important role, since it 
provides access to a population of large dust grains that responds primarily
to the gravitational dynamics of the system, rather than the radiation forces
that sculpt the small grains that dominate optical and infrared observations 
\citep{wya06}.  Multiwavelength observations of nearby systems have begun to 
reveal wavelength-dependent structure including extended haloes of small grains 
\citep[e.g.][]{su09}, often apparently generated by a much more radially 
confined ring of large parent bodies \citep{wil11,wil12}.  

HD 61005 (also known as ``The Moth'') is part of an intruiguing sample of 
debris disks with swept-back features believed to result from interactions 
with the ambient interstellar medium \citep[the others are HD 32297 and HD 
15115;][]{kal05,deb08,deb09,rod12}.  The disk has been imaged in scattered light 
both from the ground and in space, revealing a thin ring with a possible 
position offset from the central star and a symmetric pair of streamers 
originating from the ring ansae \citep{hin07,man09,bue10}.  Several mechanisms 
have been proposed to explain the unusual morphology of these swept-back 
debris disks, including two distinct ISM-driven mechanisms that result in 
similar morphology despite the orthogonal directions they assume for 
stellar motion relative to the ambient medium \citep{man09,deb09}.  By 
contrast, \citet{mar11} determine that the effects of the ISM are extremely 
sensitive to geometrical optical depth (which is directly proportional to the 
collision rate) and grain size, so that only the morphology of disks with low 
optical depths imaged at short wavelengths should be noticeably affected by 
interactions with the ISM.  If the morphology persists for high optical depths 
or large grain sizes, then alternative mechanisms such as dynamical sculpting
by an embedded planet should be considered.  It is therefore desirable to
image debris disks with swept-back structures at millimeter wavelengths, in 
order to determine the morphology of large grains in the system and thereby 
investigate the physical origins of the observed structure.  

HD 61005 presents an attractive target for millimeter-wavelength imaging.  
Located in the local bubble \citep{fra90} at a distance of 34.5\,pc 
\citep{per97} and exhibiting a ring radius of 61\,AU \citep{hin07,bue10}, it 
spans spatial scales that are amenable to being resolved with an interferometer.
It has the largest 24\,$\mu$m excess of all main sequence stars observed in the 
FEPS (Formation and Evolution of Planetary Systems) legacy survey on the {\it 
Spitzer} Space Telescope \citep{mey08}, and a bright 350\,$\mu$m flux of 
95$\pm$12\,mJy \citep{roc09}, which predicts a substantial millimeter-wavelength
flux for imaging.  As a Solar analogue \citep[spectral type G 3/5 V;][]{hil08} 
it is of particular interest for understanding the history of our own solar 
system at younger ages.  Stellar age estimates have varied around 100\,Myr, but 
recent work suggests that it is likely a member of the Argus association 
\citep{des11}.  Membership in the Argus association would lower its age to 
40\,Myr, which is more consistent with its large 24\,$\mu$m excess.

Here we use interferometric millimeter-wavelength imaging with the Submillimeter
Array (SMA) to investigate the wavelength-dependent structure of the disk 
around HD 61005.  In Section \ref{sec:obs}, we present SMA observations at a 
wavelength of 1.3\,mm that resolve the structure of the large grain populations 
for the first time.  We analyze the spectral energy distribution (SED) and 
interferometric visibilities simultaneously to study the spatial distribution 
and thermal properties of the grains in Section~\ref{sec:analysis}.  In Section 
\ref{sec:discussion} we discuss the implications of our observations in the 
context of proposed mechanisms for creating swept-back structures.  In Section 
\ref{sec:conclusions} we summarize the main results of our investigation and 
emphasize the value of future observations for providing insight into the 
underlying physical processes shaping the system.  

\section{Observations}
\label{sec:obs}

HD~61005 was observed with the SMA for four nights between 2008 December and
2012 January.  The basic observational parameters are listed in
Table~\ref{tab:obs}.  The first three nights of observation were carried out
in the SMA's compact configuration, while the final night was conducted in
the extended configuration (see table for baseline lengths).  Due to the 
southern declination of the source, the spatial resolution was generally finer
in the east-west direction than the north-south direction.  The weather was 
generally very good, with low 225\,GHz opacity, indicating high atmospheric 
transparency to millimeter-wavelength radiation, and stable atmospheric phase.  
The quasar J0747-331, located only 2.6$^\circ$ away from HD 61005, was used as 
the gain calibrator for all four tracks to correct for atmospheric and 
instrumental variations in amplitude and phase; the derived flux for the quasar 
is listed in the final column of Table~\ref{tab:obs}.  Observations of the gain 
calibrator and source were interleaved with the quasar J0826-225, which was 
used to test the efficacy of the phase transfer.  Bandpass calibration was 
carried out using a bright quasar (3c84, 3c273, or 3c454.3).  The absolute flux
scale was set using solar system objects (specified for each track in 
Table~\ref{tab:obs}); we assume a standard (but conservative) 20\% systematic
flux uncertainty due to uncertainties in the flux models for these objects.  
The correlator was configured to maximize continuum sensitivity by utilizing 
the largest available bandwidth.  The total bandwidth in the 2008 track is 
2\,GHz per sideband, with a sideband separation of $\pm$5\,GHz from the local 
oscillator (LO) frequency listed in the table. The bandwidth was upgraded to 
4\,GHz per sideband for the three subsequent tracks.  Routine calibration tasks 
were carried out using the MIR software package\footnote{See http://cfa-www.harvard.edu/$\sim$cqi/mircook.html.},
and imaging and deconvolution were accomplished with MIRIAD.

\begin{table*}
\caption{SMA Observational Parameters}
\resizebox{\textwidth}{!}{
\begin{tabular}{cccccccccc}
\hline
Date & Antennas & Baselines & $\tau_\mathrm{225 GHz}$ & LO Freq & RMS noise$^a$ & Synthesized Beam$^a$ & Flux Cal & Gain Cal & Derived Flux \\
 & & (m) & & (GHz) & (mJy\,beam$^{-1}$) & (arcsec) & & & (Jy) \\
\hline
2008 Dec 16 & 7 & 16-68 & 0.10-0.15 & 225.497 & 1.1 & 6.5$\times$2.8 & Uranus & J0747-331 & 0.90 \\
2009 Dec 29 & 8 & 16-77 & 0.06 & 225.169 & 0.6 & 5.7$\times$2.9 & Uranus/Titan & J0747-331 & 0.80 \\
2010 Apr 13 & 8 & 16-69 & 0.04 & 225.169 & 0.8 & 6.2$\times$3.0 & Titan & J0747-331 & 1.02 \\
2012 Jan 29 & 7 & 50-226 & 0.05-0.09 & 225.497 & 0.7 & 2.1$\times$1.9 & Uranus & J0747-331 & 0.95 \\
\hline
\end{tabular}
}
\tablerefs{a~Naturally weighted image}
\label{tab:obs}
\end{table*}

%Results
\section{Results}
\label{sec:results}

We detect the disk around HD 61005 independently on all four nights.  The 
combined map and visibilities are displayed in Figure \ref{plot:Image}.  The 
disk is strongly detected, with a peak signal-to-noise ratio of $\sim$8 in two
separate beams.  An appropriate shift to the visibilities has been applied to 
each data set to account for the proper motion of the star \citep[-56 and 
75\,mas\,yr$^{-1}$ in $\alpha$ and $\delta$, respectively;][]{lee07}.  The 
centroid of the millimeter-wavelength emission is consistent to within the uncertainties with the expected 
J2000 position of the star ($\alpha$ = $7^{h}35^{m}47.462^{s}$, $\delta$ = $-32^{\circ}12^{\prime}14.043^{\prime\prime}$).  

The IR excess from the literature suggests that the 
disk is optically thin, and the double-peaked structure we observe is 
consistent with the expected morphology for a highly-inclined disk with a 
central cavity.  While the structure is clearly not well described by a 
Gaussian, we estimate some basic structural parameters by fitting an 
elliptical Gaussian to the visibilities using the MIRIAD task \verb&uvfit&.  
The fit yields a major axis of 4\farcs4$\pm$0\farcs6 (150$\pm$20\,AU) and a 
minor axis of 0\farcs03$\pm$0\farcs6 (0.1$\pm$20\,AU), implying that the disk 
is spatially resolved along the major axis but not along the minor axis.  The 
position angle is $71^\circ \pm 5^\circ$, consistent with the value obtained 
from scattered light imaging, $70.3^\circ$ with a $\sim$$1^{\circ}$ uncertainty 
\citep{bue10}.  The integrated flux estimated from the elliptical Gaussian 
fit is 7.9$\pm$0.7\,mJy, although a far more reliable value will be obtained 
in the MCMC analysis in Section~\ref{sec:MCMC}.  \citet{bue10} also report 
that the center of the ring is offset from the star along the major axis by 
2.75$\pm$0.85 AU toward the SW peak.  We measure the offset of the center 
of the Gaussian compared to the star's position, corrected for proper motion.  
The centroid of the Gaussian is consistent with the star position to within 
$\pm$7\,AU; the uncertainties are too large to confirm the offset observed 
in scattered light by \citet{bue10}.  

\section{Analysis}
\label{sec:analysis}

To determine both the geometric properties of the disk and thermal properties 
of the constituent dust we simultaneously model the Spectral Energy 
Distribution (SED) and the resolved 1.3\,mm visibilities 
(Section~\ref{sec:sed}).  The best fit models are obtained from Markov Chain 
Monte Carlo fitting (Section~\ref{sec:MCMC}).  We begin by fitting a 
geometrically thin ring, motivated by the observed scattered light morphology.  
To place the results of MCMC fitting in context, we then investigate the 
effects of the assumed dust properties (Section~\ref{sec:temp}) and ring width
(Section~\ref{Sec:Width}).  We also conduct a visibility-domain comparison of
the millimeter-wavelength emission with the known scattered light morphology to
investigate the relative spatial distribution of small and large grains
(Section~\ref{sec:wings}). 

\subsection{Disk Model}
\label{sec:sed}

\begin{figure}[ht]
\begin{center}
\epsscale{1}
\plotone{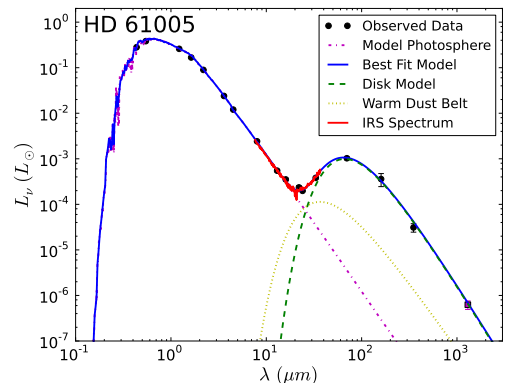}
\caption{SED of the HD 61005 system.  The total model SED is the sum of three 
components:  a Kurucz-Lejeune model photosphere, the debris disk, and a warm 
100K dust belt.  The flux we measure at 1.3\,mm, which is calculated from our 
best fit model and is not included in the calculation of the SED $\chi^2$ 
value, is displayed with a magenta square.  The error bars for this point 
include an estimated 20\% systematic uncertainty.  The units are defined as 
$L_\nu = 4 \pi d^2 \nu F_\nu$ in units of $L_\odot$.  Our model coincides well 
with the Spitzer IRS spectrum, despite the fact that these data were not used 
in the fitting process and are displayed for comparison only.
  \label{plot:SED_mcmc1119}}
\end{center}
\end{figure}

As shown in Figure \ref{plot:SED_mcmc1119}, we model the SED with three 
components:  (1) a Kurucz-Lejeune model photosphere with surface gravity log 
$g$ = 4.5, effective temperature $T_{eff}$ = 5500 K, and solar metallicity 
Z = 0.01 \citep{des11}, (2) a cold, spatially-resolved outer debris disk, and 
(3) a warm inner dust belt modeled as a single-temperature blackbody.  The 
addition of the belt is necessary to increase the flux around 20 microns, where 
neither the star nor the disk contribute enough flux to account for observations.  
The short-wavelength excess that we attribute to the presence of an inner belt 
could equally arise from hot emission from a population of grains smaller than the 
characteristic grain size in our model; however, the data at this point are 
insufficient to distinguish between the two scenarios.  Accordingly, the 
properties of the warm belt are not well constrained, since it only produces 
a substantial contribution to the total flux over very small portion of the 
observed range of wavelengths, so for simplicity we allow only the mass of 
the belt to vary, and parameterize it as a narrow ring of 100\,K dust (we also 
demonstrate below that our results are not sensitive to the assumed 
temperature).  The cold disk, which we model as a spatially extended component, 
contributes essentially all of the flux at 1.3\,mm; in our best fit, the 
warm dust belt contributes only 0.36 mJy to the total flux at this wavelength, a 
factor of 20 less than that of the extended component.  This modeling 
procedure is similar to that presented in \citet{hug11}, with the primary difference
being the use of astrosilicate opacities to determine grain temperatures.  We 
briefly describe the salient features of the model below.  

%SED

Each disk model is determined by 6 free parameters:  the inner radius of 
the disk ($R_{in}$), the characteristic grain size ($a$), the disk mass 
($M_D$), the grain emissivity parameter ($\beta$), the belt mass ($M_B$), 
and the width of the disk ($w$).  In our initial fitting efforts, we fix the
width of the disk $w$ to a small, constant fraction of 5\% of $R_{in}$ to 
match the scattered light morphology (although this assumption is relaxed 
in Section \ref{Sec:Width} below).  $R_{in}$ is mostly constrained by the 
visibilities, but it also has an effect on the equilibrium temperature of the 
dust grains, which directly affects the peak flux and wavelength of the SED.  
The grain size, $a$, determines the temperature of the grains.  In reality 
the disk most likely has a broad distribution of grain sizes, however assumption of a single characteristic grain size is sufficient to reproduce the observed SED while minimizing 
computational requirements. 
%(although as previously mentioned it might 
%alternatively be possible to reproduce the observed 20\,$\mu$m 
%excess with a population of grains significantly smaller than the 
%characteristic grain size).  
$M_D$ is essentially a luminosity scaling factor for the disk SED, just as 
$M_B$ scales the flux of the belt.  Finally, $\beta$ determines the slope 
of the Rayleigh-Jeans tail.  In accordance with \citet{wil04}, we assume a 
dust grain emission efficiency $Q_\lambda = 1 - \exp[-(\lambda/\lambda_0)^{-\beta}]$, 
where $\lambda_0 = 2 \pi a$ is the critical wavelength.  This function has 
the desired asymptotic properties that $Q_\lambda \approx (\lambda 
/ \lambda_0)^{-\beta}$ when $ \lambda >> \lambda_0$, and $Q_\lambda 
\approx 1$ when $\lambda << \lambda_0$. This parametrization of $Q$ 
does {\it not} factor into our calculation of the grains' temperature, however.  
It is required to ensure smooth long-wavelength emission in the SED by 
mimicking the effect of a grain size distribution, since grains at the characteristic grain size are extremely inefficient emitters at $\lambda >> a$.  

In order to determine the temperature of a grain of a given size and distance from the star,
we assume that the dust composition is compact astrosilicates \citep{dra03} and obtain the grains' opacity, $\kappa_{tot} (a,\lambda)$, and albedo, $\omega (a,\lambda)$, 
using Mie theory \citep[see, e.g.][]{boh83} as implemented in the radiative transfer code MCFOST \citep[][]{pin06,pin09}.  We then derive the appropriate grain temperature from energy balance.  As written in \citet{tie05}, the energy emitted per unit time from the grain can be expressed:
\begin{equation}
\Gamma_{out} = 4 \pi \cdot \pi a^2 \cdot \int_{0}^{\infty} Q(a,\lambda) B(T,\lambda) d\lambda
\end{equation}
where $Q(a,\lambda)$ represents the fraction of emission at wavelength $\lambda$, and $B(T,\lambda)$ is the grains' Planck function.  Similarly, the power absorbed from the star can be written:
\begin{equation}
\Gamma_{in} = \pi a^2 \cdot \int_0^{\infty} Q(a,\lambda)F_{\lambda}(r,\lambda) d\lambda
\end{equation}
where $F_{\lambda}$ is determined from our Kurucz-Lejeune model of the stellar photosphere, $a$ is the grain size, and $r$ is the distance from the star to the grain.  
Rather than using the simplistic parametrization of $Q$ mentioned above to determine the flux density of the Rayleigh-Jeans tail, we obtain $Q(a,\lambda)$ from $\kappa_{tot} (a,\lambda)$ and $\omega (a,\lambda)$.  Assuming spherical grains, the absorption efficiency is:
\begin{equation}
Q(a,\lambda) = \frac{4}{3} \kappa_{tot}(a,\lambda) \rho a (1-\omega(a,\lambda))
\end{equation}
where $\rho$ is the mass density of a single grain.%, which is assumed to equal $2.7$\,g\,cm$^{-3}$.  
Setting $\Gamma_{in} = \Gamma_{out}$, we arrive at:
\begin{equation}
\frac{1}{4\pi} \int_0^{\infty} Q(a,\lambda)F_{\lambda}(r,\lambda) d\lambda = \int_{0}^{\infty} Q(a,\lambda)B(T,\lambda)d\lambda \label{Temperature_Integral}
\end{equation}
By numerically evaluating each integral in Equation~\ref{Temperature_Integral}, we obtain the temperature of a given grain.  In order to do so, we generate a lookup table of these integrals.  The integral on the left is tabulated as a function of $a$ and calculated at a radius of 50 AU; to determine the values of this integral at different radii, we scale these values by $1/r^2$.  The second integral is tabulated as a function of both $a$ and $T$.  $a$ is sampled from $0.1$ to $3000$ microns with about 3 sizes per decade (in log space), and $T$ is sampled from $0$ to $1000$ K with a step size of 1 K.  The associated wavelengths in our lists of $\kappa_{tot}(a,\lambda)$ and $\omega(a,\lambda)$ span from $10^{-5}$ to $3\times10^{-1}$\,cm, sampled in logarithmic intervals of 0.05.  

The surface number density of grains, $N(r)$, is related to the 
surface mass density as $\Sigma (r) = N(r) m_g$, where $m_g$ is the mass of a 
grain, which we assume is spherical.  We assume a density of 2.7\,g\,cm$^{-3}$, which is a 
compromise between typical bulk densities measured for cometary and 
interplanetary dust particles and terrestrial materials typically assumed to 
comprise astronomical graphite or silicate grains \citep[see, e.g.,][]{dra84,
bro06,blu08}.  We then parameterize 
$\Sigma (r) = \Sigma_{100}(\frac{r}{100 \mathrm{ AU}})^{-p}$, where $p$ is the 
surface density power law, which we fix at  a value of 1 (this value is consistent 
with both the radial falloff of surface brightness in the region of the AU Mic birth 
ring postulated as the birth ring in \citet{str06}, as well as a typical value for 
bright protoplanetary disks measured by \citet{and09}).  There is a well-known 
degeneracy between $p$ and the outer radius, and our data are not of 
sufficiently high quality to distinguish between these variables; see, e.g., 
discussion in \citet{mun96}.  We integrate flux contributions between the inner and 
outer radius, yielding:
\begin{equation}
F_\lambda =  \frac{\pi a^2 Q(\lambda)}{d^2} \int_{R_{in}}^{R_{out}} 2 \pi r B_\lambda (T_r) N(r) dr
\end{equation}

At the observed wavelength of 1.3\,mm, we use the equations above to generate a 
high-resolution synthetic image of the disk.  As is evident in Figure 
\ref{plot:SED_mcmc1119}, the star and the 100\,K dust 
belt do not contribute significant flux at the wavelength of the SMA 
observations, and are hence absent from the model image.  To generate this 
image, we assume an inclination angle of $84.3^\circ$ and a position angle of 
$70.3^\circ$ derived from the higher-resolution scattered light images 
\citep{bue10}, and project the flux onto the sky plane.  We then utilize the 
MIRIAD task \verb&uvmodel& in order to sample the model image at the same 
spatial frequencies as the SMA data and compare the data and model in the 
visibility domain.

\subsection{MCMC Fitting \label{sec:MCMC}}

In order to compare the model with the SED and visibilities, we compute a 
$\chi^2$ value for each and sum the two.  As discussed in \citet{and09}, these 
two values of $\chi^2$ are comparably sensitive to changes in the disk 
parameters (the large numbers of visibilities are balanced by the low fractional
uncertainty on the SED points), causing neither to dominate the total $\chi^2$ 
and thus the final fit.  We omit the SMA 1300\,$\mu$m flux from the SED $\chi^2$ 
calculation since it is implicitly included in the visibility $\chi^2$.  In 
order to locate the best fit and determine constraints on each parameter, we 
utilize the Markov Chain Monte Carlo (MCMC) fitting technique described in 
\citet{goo10}, an affine-invariant ensemble sampler which performs well when 
parameters are correlated.  The SED includes the observed fluxes listed in 
Table \ref{table:Fluxes}.  Only the points with wavelengths beyond 10 $\mu$m 
are affected by the parameters of the model disk, and therefore we include 
only these values in the computation of $\chi^2$.

%Observed Fluxes
\begin{table} [ht]
\begin{center}
\caption{Observed Flux Points for HD 61005\label{table:Fluxes}}
    \begin{tabular}{ ccc p{6cm} }
    \hline
    $\lambda$ ($\mu m$) & Flux (Jy) & Source\\ \hline
    0.436 & $1.04 \pm 0.02$ & Tycho-2 \citep{hog00} \\
    0.545 & $1.82 \pm 0.02$ & Tycho-2 \citep{hog00} \\
    1.220 & $2.77 \pm 0.07$ & 2MASS \citep{cut03} \\
    1.630 & $2.39 \pm 0.10$ & 2MASS \citep{cut03} \\
    2.190 & $1.71 \pm 0.04$ & 2MASS \citep{cut03} \\
    3.6 & $0.75 \pm 0.02$ & FEPS \citep{hil08} \\
    4.5 & $0.47 \pm 0.01$ & FEPS \citep{hil08} \\
    8 & $0.17 \pm 0.004$ & FEPS \citep{hil08} \\
    13 & $0.062 \pm 0.004$ & FEPS \citep{hil08} \\
    16 & $0.049 \pm 0.003$ & Spitzer IRS Archive \\
    22 & $0.045 \pm 0.004$ & Spitzer IRS Archive \\
    24 & $0.041 \pm 0.002$ & FEPS \citep{hil08} \\
    33 & $0.11 \pm 0.007$ & FEPS \citep{hil08} \\
    70 & $0.63 \pm 0.05$ & FEPS \citep{hil08} \\
    160 & $0.50 \pm 0.16$ & FEPS \citep{hil08} \\
    350 & $0.095 \pm .012$ & CSO \citep{roc09} \\
    \hline
    \end{tabular}
\end{center}
\end{table}

After experimentation with the initial values of the ensemble, we find that 
the ensemble consistently converges to the same region in the parameter space 
independent of the initial parameter values.  For all subsequent runs, we fix 
the initial values of the chain to the values of a reasonably good fit:  
$R_{in} = 68$ AU, $\log(a [\mu m]) = 0.5$,  $\log(M_D [M_\oplus]) = -2.7$, 
$\beta = 0.5$, $\log(M_B [M_\oplus]) = -6.0$.  
Trial states for $R_{in}$ and $\beta$ are generated in linear space, while 
states for $a$, $M_D$, and $M_B$ are generated in logarithmic space.  The 
widths of the Gaussians determining trial states for each parameter are set 
to 2 AU for $R_{in}$, 0.05 for log($a$), 0.1 for log($M_D$), 0.01 for $\beta$, 
and 0.05 for log($M_B$), which we found were the approximate uncertainties in these parameters. We run $100$ ``walkers'' through $800$ trials each, and 
after rejecting the ``burn-in" phase (the region in which the average $\chi ^2$ 
decreases with time before settling) which constitutes the first $\sim 200$ trials, 
we then determine the best-fit model and generate probability 
distributions for each of the parameters.   

\begin{figure*}[ht]
\begin{center}
\epsscale{1.0}
\plotone{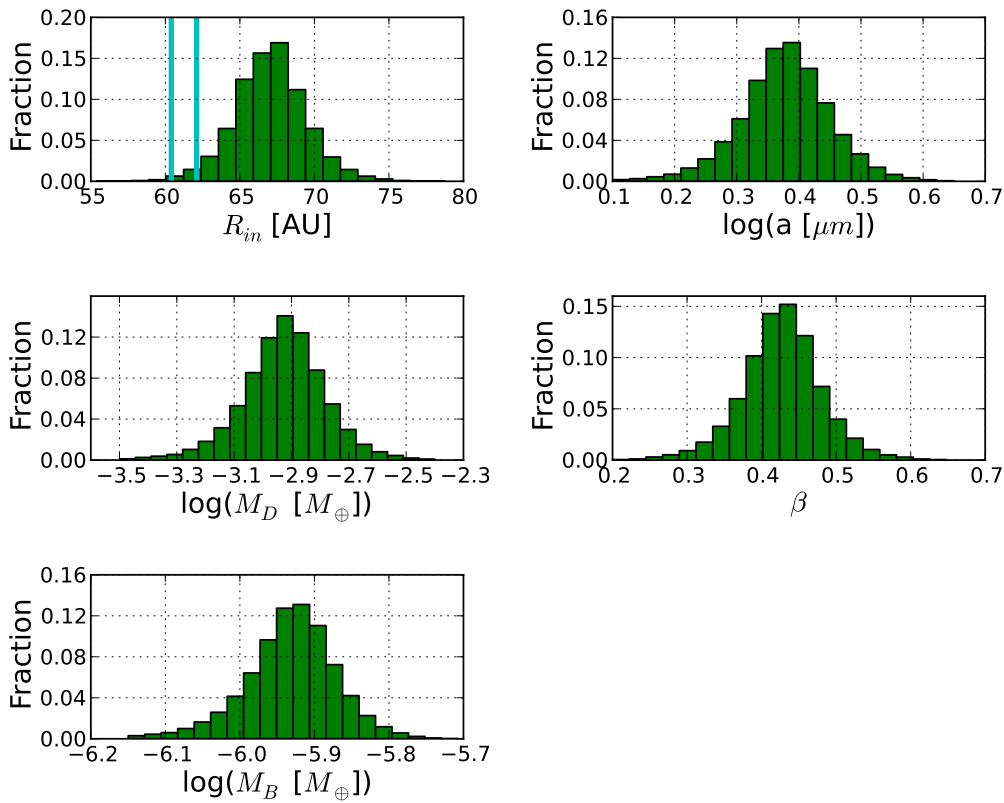}
\caption{Probability distributions for the model parameters as derived from the 
MCMC chain in the case of a narrow ring ($w/R = 5\%$).  The height of 
each bar represents the fraction of the models that fall into its respective 
bin.  In the $R_{in}$ plot,
the cyan vertical lines represent the $\pm 1 \sigma$ range in radius derived in \citet{bue10}.  
  \label{plot:histo_mcmc1119.eps}}
\end{center}
\end{figure*}

\begin{figure*}[ht]
\begin{center}$
\begin{array}{cc}
\hspace{-35pt}
\includegraphics[width=0.4\textwidth]{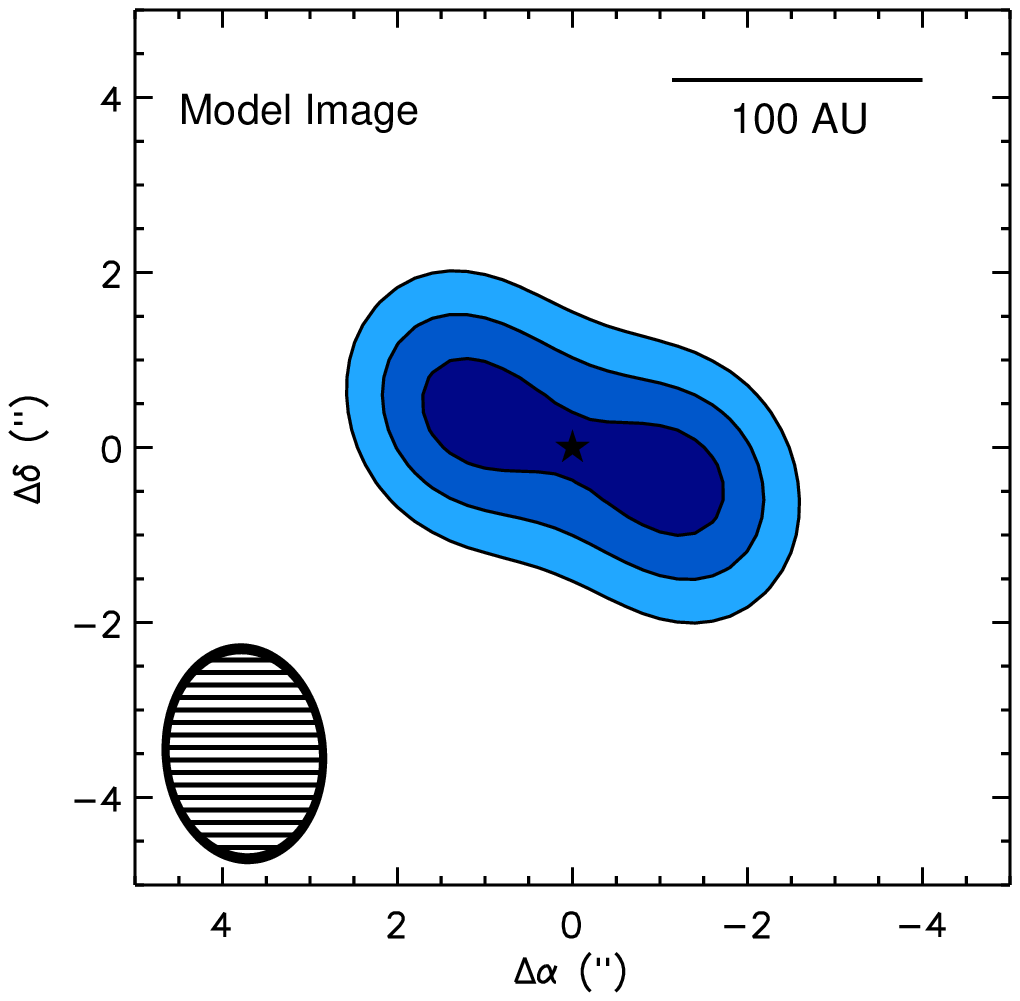} &
\includegraphics[width=0.4\textwidth]{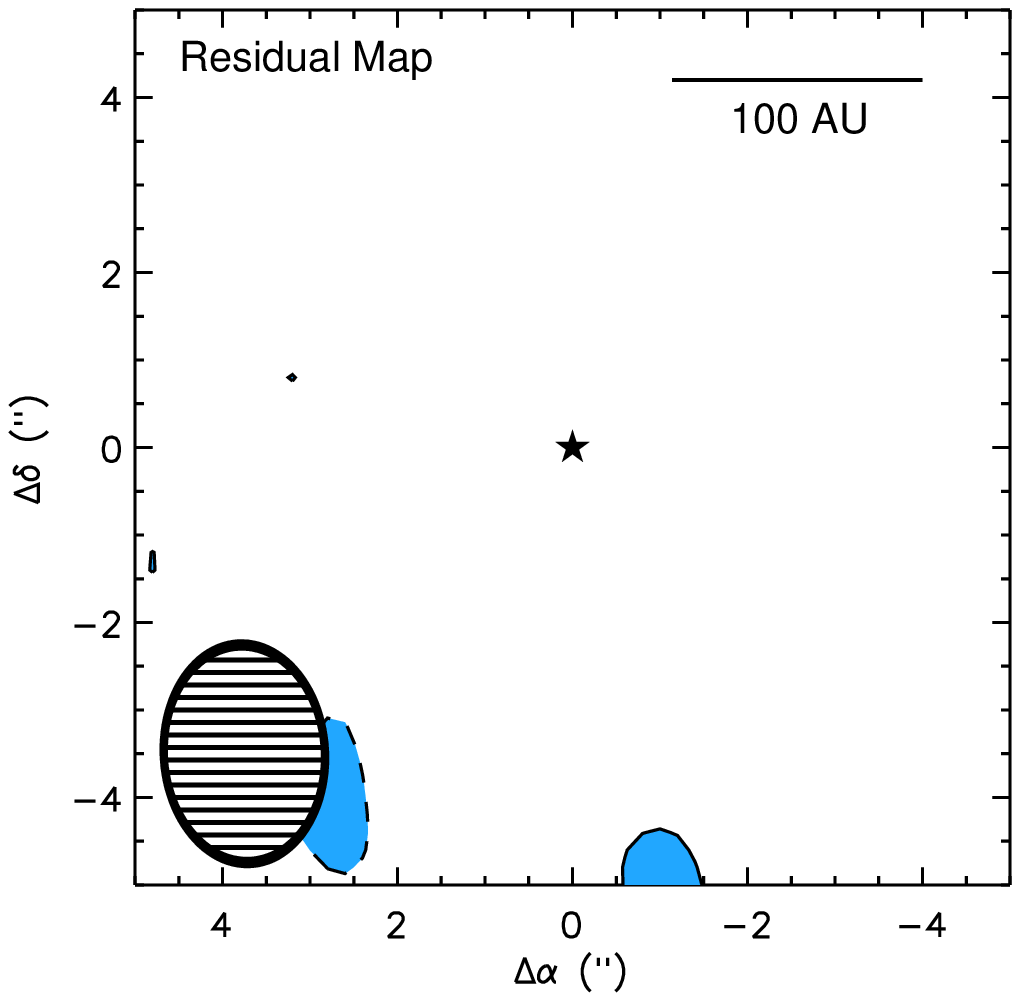}
\end{array}$
\vspace{-30pt}
\caption{The best fit observed model image and corresponding residual map.  
For comparison with Figure~\ref{plot:Image}, [3,5,7] $\times$ 0.34 mJy contours are drawn in the observed model image.  In the residual 
map, the $+2\sigma$ contours are drawn with solid lines, while the $-2\sigma$ 
contours are drawn with dashed lines.   The residual emission does not exceed the 3$\sigma$ level.
\label{plot:Model_Image.eps}}
\end{center}
\end{figure*}

Best fit parameters for the five free parameters are listed in the left column 
of Table \ref{table:BestFitParameters}.   Figure 
\ref{plot:histo_mcmc1119.eps} displays probability density functions 
generated by the models of the chain, as well as the locations of the best fit 
parameters, defined as the mode of the probability distributions.  Figure 
\ref{plot:Model_Image.eps} displays the best fit model image, as well as its 
residuals.  For this best fit, we obtain a total flux of $7.2 \pm 0.3$ mJy 
(with an additional estimated 20\% systematic 
uncertainty).  For this value, and the values in Table \ref{table:BestFitParameters}, the uncertainty is determined from the posterior PDF of the ensemble as the width in parameter space which encloses 68.2\% (1 $\sigma$) of the models, which we found to be symmetric about the best fit.  For each of these models, the reduced $\chi^2$ is equal to $1.85$.

%Histograms

%Model Image

%Best Fit Parameters
\begin{table} [h!]
\begin{center}
\caption{The best fit parameters for HD 61005. \label{table:BestFitParameters}}
    \begin{tabular}{ cccc }
    \hline
    Parameter 
	& Best-Fit Model 
	& For $T_B = 80$\,K 
	& For $w = 65$\,AU \\ \hline
    $R_{in}$ (AU) 
	& 67 $\pm$ 2
	& 67$\pm$ 2
	& 71$^a$ $\pm$ 3 \\
    $\log(a\ [\mu \rm{m}])$ % ($log(\mu m)$) 
	& 0.38 $\pm$ 0.07 
	& 0.44 $\pm$ 0.09
	& 0.33 $\pm$ 0.09 \\
    $\log(M_D\ [M_\oplus])$ % ($log(M_\oplus)$) 
	& $-2.92 \pm 0.13$ 
	& $-2.92 \pm 0.17$
	& $-2.91 \pm 0.16$ \\
    $\beta$ 
	& 0.43 $\pm$ 0.05 
	& 0.40 $\pm$ 0.06
	& 0.41 $\pm$ 0.06\\
    $\log(M_B\ [M_\oplus])$ % ($log(M_\oplus)$) 
	& $-5.93 \pm 0.06$
	& $-5.23 \pm 0.06$
	& $-5.97 \pm 0.08$ \\ 
    Total $\chi^2$
	& $456103.228$
	& $456107.180$
	& $456107.413$ \\ \hline
    \end{tabular}
\tablerefs{a~This is the radius which encloses half of the total flux.}
\end{center}
\end{table}

\begin{figure*}[ht]
\begin{center}$
\begin{array}{ccc}
\hspace{-20pt}
\includegraphics[width=0.37\textwidth]{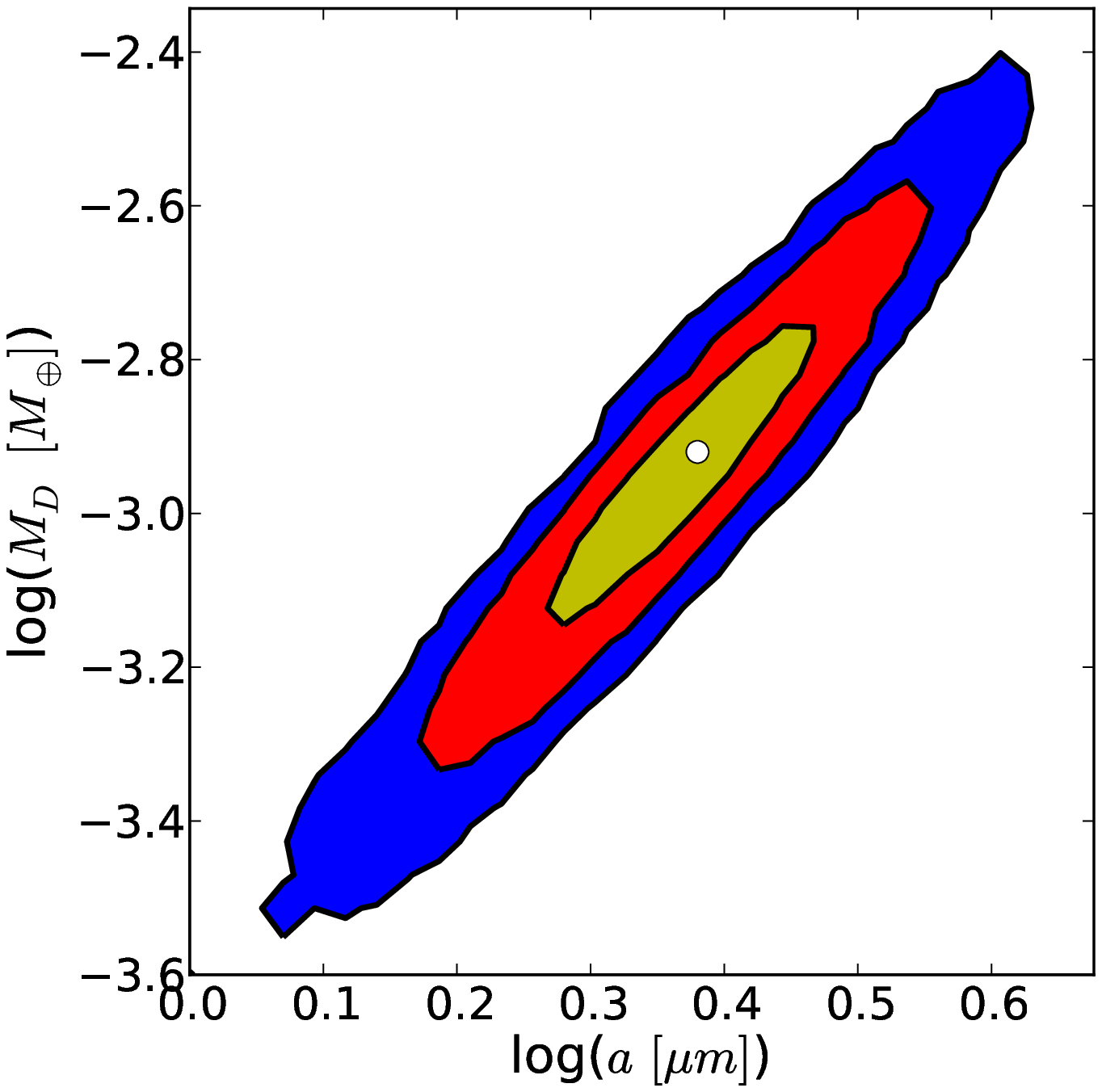} &
\hspace{-25pt}
\includegraphics[width=0.37\textwidth]{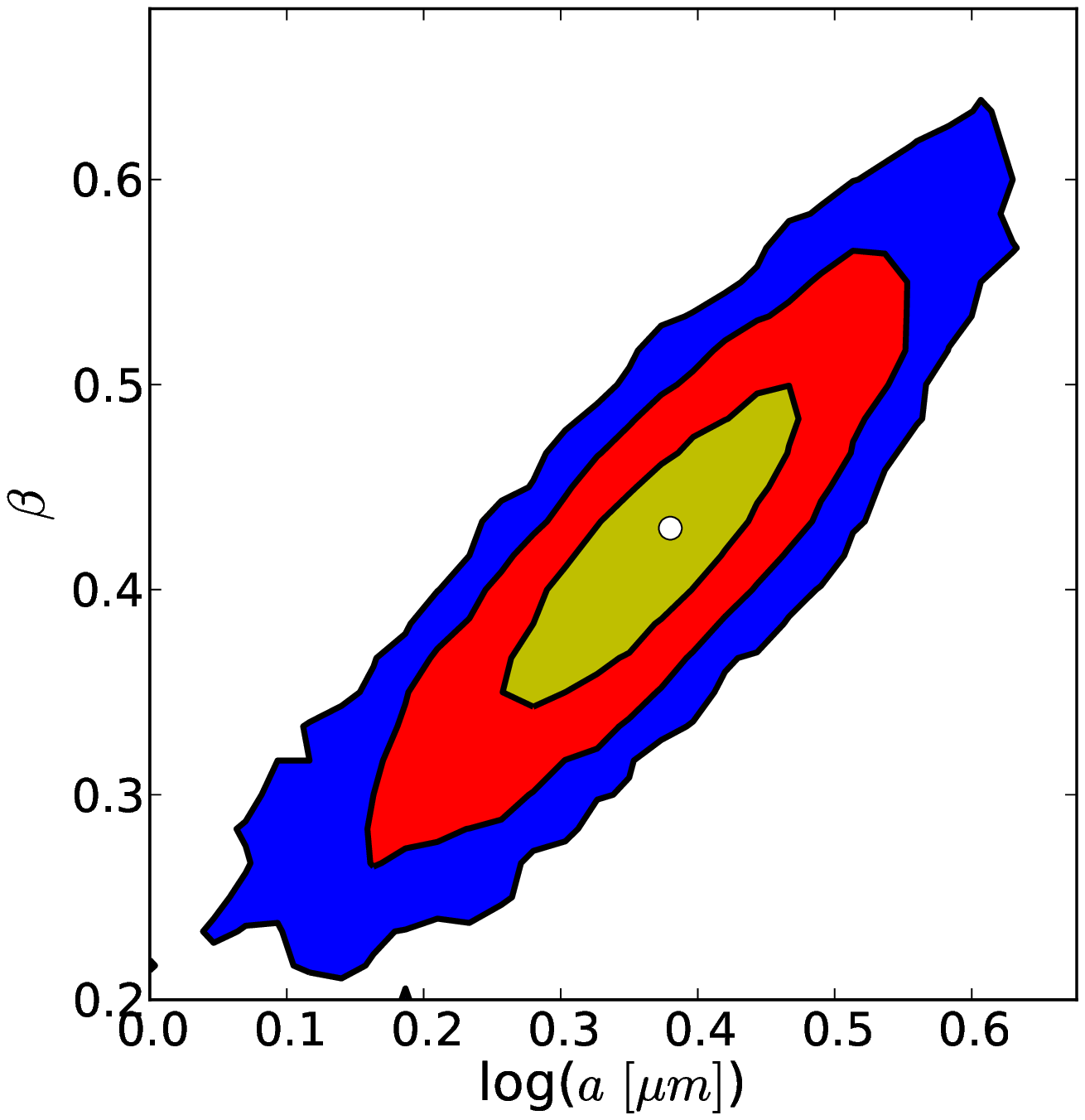} &
\hspace{-25pt}
\includegraphics[width=0.37\textwidth]{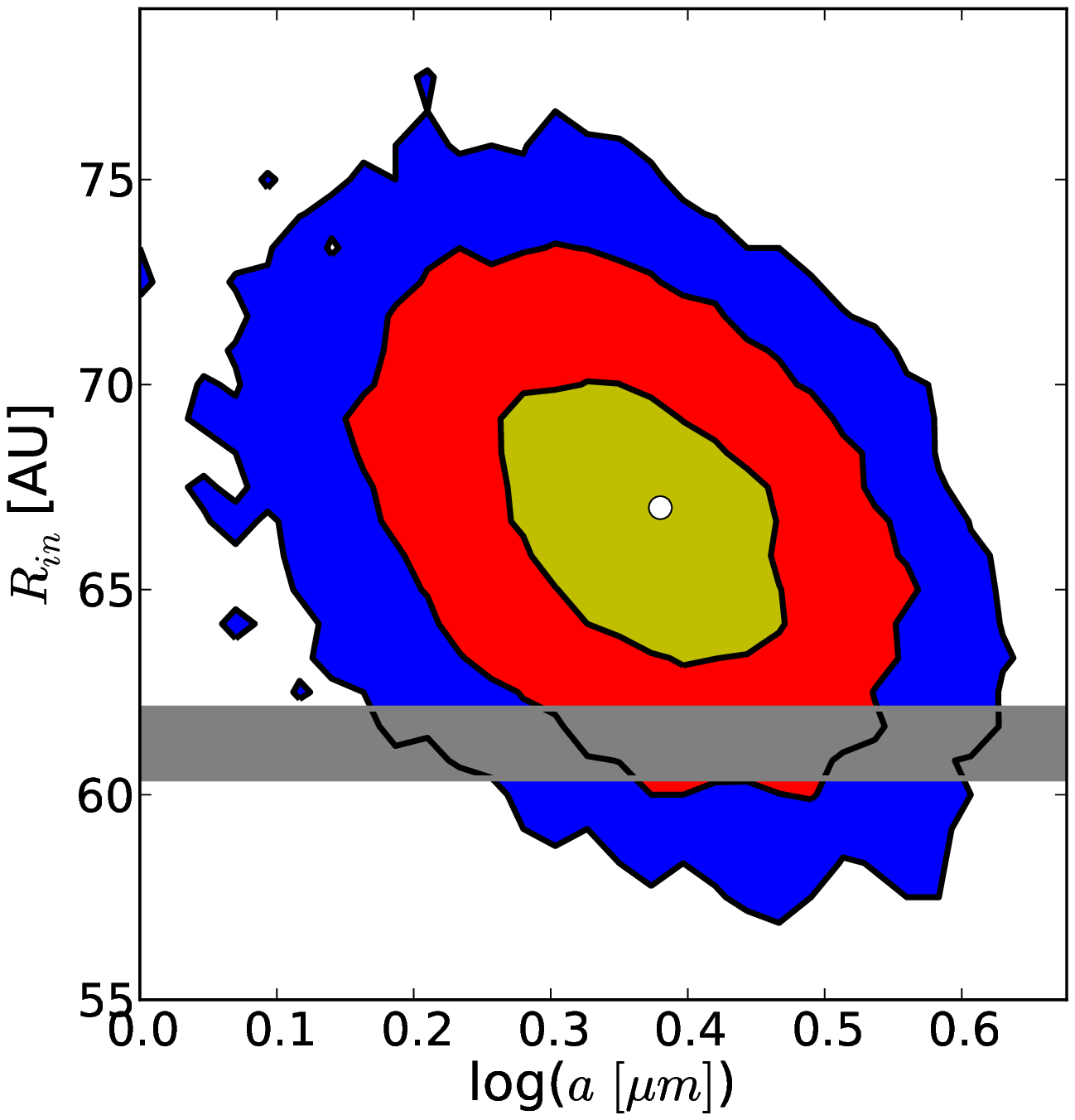}
\end{array}$
\vspace{-10pt}
\caption{The probability distribution of an MCMC chain in the 
log($a$)-log($M_D$), log($a$)-$\beta$, and the log($a$)-$R_{in}$ planes.  
Contours enclose 68.2\% (1$\sigma$), 95.4\% (2$\sigma$), and 99.6\% (3$\sigma$)
of the models, and the white circle marks the position of the global minimum $\chi^2$.
Here, the degeneracy between $a$, $M_D$, and $\beta$ is readily apparent; we
also note that the degeneracy with $R_{in}$ is broken with the use of the 
spatially resolved visibilities.  The grey rectangle in this final plot depicts
the value of $R_{in}$ determined from scattered light in \citet{bue10}.  The 
best-fit radius using the millimeter data is consistent with the scattered light radius 
to within the uncertainties.
  \label{plot:Contour}}
\end{center}
\end{figure*}

%In addition, Figure \ref{plot:Contour} displays two-dimensional views of the 
%$\chi ^2$ parameter space, revealing degeneracies between $a$, $M_D$, and 
%$\beta$.  {\bf [We should decide whether or not to leave these plots in...]}
There are known degeneracies in this fitting process.  Since we parameterized 
the disk with a single characteristic grain size, for a given value of $M_D$, increasing $a$ 
decreases the number of blackbody emitters and decreases their temperature, 
which therefore decreases the total flux.  Hence, $M_D$ must increase in 
accordance with $a$ in order to maintain the observed flux.  $\beta$ is less 
strongly correlated with $a$, but the degeneracy occurs because as $a$ 
increases, the peak wavelength in the SED also increases as the grain 
temperature decreases.  Since $\beta$ affects the slope of the Rayleigh-Jeans 
tail of the SED, it must increase along with the position of the peak in order 
to obtain a steeper slope.  If we were fitting for the SED alone, we would 
also expect $R_{in}$ to exhibit a similarly strong degeneracy, since higher 
temperatures and therefore fluxes could be obtained both by bringing the 
grains closer to the star and by shrinking the grains.  However, the inner 
radius is well-constrained by our spatially resolved data, so there is only a
mild correlation between $R_{in}$ and $a$, as shown in Figure \ref{plot:Contour}.

It is also necessary to check
that the underlying assumptions of our model do not bias these results.  In 
Sections \ref{sec:temp} and \ref{Sec:Width} we investigate the robustness of 
the best-fit results to perturbations in our assumptions about belt temperature 
and ring width.  Finally, it is important to consider our results in the 
context of the scattered light morphology.  The ring radius we derive is 
consistent with the radius reported from scattered light observations \citep[61.25$\pm$0.85\,AU;][]{bue10}, differing by 2.6\,$\sigma$.  This result hints at 
wavelength-dependent structure, in the sense that if the millimeter emission
traced the scattered light precisely, the ring radius should be noticeably 
larger to reflect the contribution from the scattered light wings.  To 
quantify the spatial distribution of the millimeter-wavelength emission, we 
perform a visibility-domain analysis of the millimeter morphology in 
Section~\ref{sec:wings}, using a toy model that incorporates constraints 
from the scattered light emission to decompose the emission into ring and 
streamer components.

\subsubsection{Effect of Assumed Belt Temperature \label{sec:temp}}

As discussed above, a simple single grain size fit to the data is incapable of
reproducing the short-wavelength flux in the IRS spectrum.  This additional 
flux necessitates either a population of hot grains substantially smaller than the
characteristic grain size (and therefore substantially smaller than the blowout
size for this system), or the addition of an inner warm dust belt.  
While there are very few data to constrain belt properties, we conducted a 
brief exploration of the effect of our assumed 100 K belt temperature on the 
derived disk parameters by running a separate MCMC chain with the temperature 
of the inner belt set to 80 K rather than 100 K.  The center column of Table 
\ref{table:BestFitParameters} displays the results of the best fit.

The only noticeable change occurs in $M_B$, which increases to compensate for 
the decreased temperature.  The visibilities fix $R_{in}$, the SED peak fixes 
$a$ and $M_D$, and the slope of the Wien tail fixes $\beta$.  This model fit
deviates in quality of fit from the 100\,K model only by 0.6$\sigma$.  The 
available data are therefore evidently not sufficient to strongly constrain the 
temperature of the warm belt, but this confirms the robustness of our analysis 
of the outer, cold debris disk, independent of the assumed belt temperature.

\subsubsection{Effect of Disk Width \label{Sec:Width}}

While the data are consistent with a narrow ring centered at the radius of 
the scattered light ring, we also investigate a scenario in which 
the width is fixed to 65\,AU to determine whether the millimeter data can 
constrain the width of the ring.  This value was chosen so that $w / R \sim 1$. 
The right column of Table \ref{table:BestFitParameters} displays the results.

The best-fit broad belt has a central radius of 71 $\pm$ 3 \,AU from the 
star (although due to the falloff of surface density with radius, more of the 
emission is concentrated closer to the star), which is consistent with the 
radius of the narrow ring fit within 2 sigma.
This model differs in quality of fit from the best-fit narrow ring model only by 
0.6$\sigma$.  Our spatial resolution -- nominally $\sim$80\,AU with natural 
weighting, but including shorter baselines that provide information on smaller 
spatial scales -- is therefore evidently not sufficient to distinguish between a 
wide and a narrow millimeter belt, although it is notable that in both the wide 
and narrow disk fit, the disk is centered on the location of the scattered light 
ring to within the uncertainties.

\subsection{Are there Large Grains in the Streamers?}
\label{sec:wings}

The swept-back ``wings" are the most remarkable features of the debris 
disk around HD 61005.  Particularly prominent in scattered light, they 
contribute slightly more than half the flux at short wavelengths.  While most theoretical
interpretations of the streamers focus on dynamics of small grains, large 
grains that dominate millimeter maps are far better tracers of gravitational 
dynamics and are therefore useful in distinguishing between mechanisms that 
might sculpt this striking morphology.  The relative contribution of the flux 
at infrared and millimeter wavelengths provides insight into the grain size 
distribution in the wings and the streamers.

Since this is a purely geometric problem and we cannot reliably distinguish 
between SED contributions from the disk and the streamers, we perform the 
analysis on the visibilities only.  We therefore depart from the analysis 
strategy of Section~\ref{sec:sed} and instead use a toy model that incorporates 
constraints on the disk structure from the observed scattered light morphology.  
Unlike the MCMC analysis in Section \ref{sec:MCMC} above, we fix as many 
parameters as possible to match the scattered light.  To maintain consistency 
with \citet{bue10}, we fix the inner radius of the disk to 61.25\,AU and the
width to the original narrow value.  We set the angle that the streamers make 
with the plane of the disk to $23^\circ$, and set the flux density power law 
to that observed in scattered light: $F \propto r^{-4.4}$ \citep{bue10}.  

\begin{figure}
\begin{center}
\epsscale{1}
\plotone{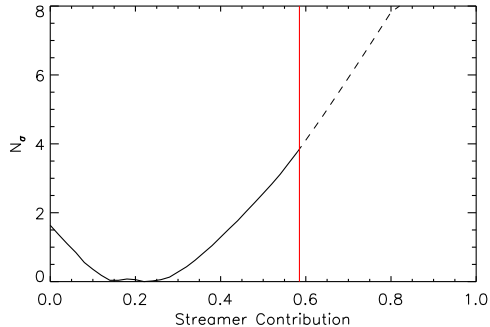}
\caption{Deviation from the best fit model as a function of the fraction of 
millimeter-wavelength emission contributed by the streamers.  The deviation is 
expressed in units of standard deviations (derived from $\Delta \chi^2$, 
assuming a Gaussian probability distribution).  This plot effectively shows 
the fraction of 1.3\,mm emission contributed by the streamers in the best-fit 
model (the minimum is near 0.2) and the level at which other flux fractions are 
ruled out in the context of our toy model.  The data are consistent with a 
scenario in which the bulk of the millimeter emission originates from the 
narrow scattered light ring.  The value corresponding to a scenario in which 
the millimeter emission traces the scattered light morphology is marked with 
a vertical red line, and is ruled out at the $\approx 4$\,$\sigma$ level.  
  \label{plot:Streamergrid}}
\end{center}
\end{figure}

We compute a grid that varies the total flux in the image and the streamer 
contribution.  For each value of the streamer contribution between zero and
one, we search for the value of the total flux between 5 and 9 mJy that 
minimizes the $\chi^2$ value between the data and model.  Each of these ranges 
of parameters was divided into 50 steps, such that the streamer contribution 
was explored with a step size of 2\%, while the total flux was explored with 
a step size of 0.08 mJy.  We then calculate the statistical deviation of each 
model from the global minimum.  Figure \ref{plot:Streamergrid} displays the 
results.  The deviation from the global minimum $\chi^2$ value, expressed as 
the number of standard deviations corresponding to the probability with which 
the model is a worse fit than the global best-fit value, is plotted against 
the percentage of the total flux in the image that originates in the streamers 
(with the remainder of the flux originating from the ring).  The observed 
streamer contribution from the scattered light data is indicated by a vertical 
red line.  The results demonstrate that a scenario in which the millimeter
emission traces the morphology of the scattered light, including the swept-back
wings, is ruled out at the 4\,$\sigma$ level.  The data are of insufficient
quality to draw fine distinctions about whether some smaller fraction of the 
millimeter flux might originate from the streamers, but they are certainly 
consistent with a scenario in which the large grains responsible for the 
1.3\,mm emission are confined exclusively to the scattered light ring.  The 
implications of this result will be discussed in more detail in Section 
\ref{sec:structure} below.    

\section{Discussion}
\label{sec:discussion}

%Contours

\subsection{Dust Grain Properties}
\label{sec:dust}

The best-fit grain size in the fiducial model is $a$=2.4\,$\mu$m.  This value is 
somewhat smaller than for other debris disks analyzed using similar methods 
\citep[see, e.g.,][]{wil04,hug11}; the relatively small grain size reflects the high
temperatures indicated by the peak wavelength of the SED, which is inconsistent
with blackbody equilibrium temperature at the radii indicated by the millimeter 
and scattered light images.  The blow-out grain size for HD 61005 due to 
radiation pressure is roughly 1$\mu$m, estimated using the relationship 
$a_\mathrm{blowout} = 3L_*/16\pi G M_* c \rho$, where $L_*$ is the stellar 
luminosity, $M_*$ is the mass of the star, $c$ is the speed of light, and $\rho$ 
is the density of a dust grain.  This relationship results from balancing the 
radiation pressure force against the gravitational force for a particle on a 
circular orbit (which is a factor of two easier to remove from the potential 
well than a stationary dust grain), including a factor of 0.5 to estimate the 
effects of a realistic albedo and radiative efficiency of a silicate dust grain 
\citep{bac93}.  

It is clear from the ensemble of models in our fitting process 
that the small grain size primarily reflects the need for the grains to attain 
a temperature high enough to match the peak wavelength of emission while 
being located in a ring roughly 60-70\,AU from the central star.  A simple 
estimate of dust temperature using Wien's Law and the approximate peak 
wavelength of the blackbody predicts a dust temperature of about 43\,K, 
while the temperature predicted by Equation \ref{Temperature_Integral} is 50\,K for 
grains with the best-fit size and emission efficiency.  
%The blackbody equilibrium temperature for grains at 
%60\,AU from this star would be nearly 20\% colder.  This discrepancy in 
%temperature is primarily a reflection of our overly simplistic model 
%assumptions, in particular the assumption of a single grain size and 
%associated emissivity rather than a more realistic grain size distribution.  
As a sanity check, we performed a ``chi-by-eye'' fit to the SED using a 
complete radiative transfer code with a realistic grain size distribution 
\citep[MCFOST;][]{pin06,pin09}.  We obtained a reasonable fit with a minimum 
grain size of 2.5\,$\mu$m.  The ring is assumed to extend from 60 to 63 AU, 
with a surface density power law index $p$=0.5 and a total mass in dust of 
$3.3\times10^{-4}$ M$_\earth$. The assumed dust composition is compact 
astrophysical silicates \citep{dra03}, with a distribution spanning the 
2.5\,$\mu$m--1\,cm range and a -3.5 power law index appropriate for a 
collisional cascade.  However, we did not pursue MCMC minimization with 
this code due to the computational intensity of the task.  We mention it here 
only to indicate that it is possible to reproduce the SED (although not the 
near-IR excess that we attribute to the warm belt) with a realistic 
grain size distribution that includes only grains larger than the blowout size. 

It is also worth noting that SED modeling has been notoriously difficult due to
the complexity of the system.  Previous SED models using single-blackbody fits
have resulted in disk radius estimates between 16 \citep{hin07} and 96\,AU 
\citep{roc09}, depending on the method and assumed grain properties, while
a slightly more complex extended disk model by \citet{hil08} predicted a disk
that stretches between radii of 8.6 and 41\,AU.  To some extent the small radii
indicate the presence of hot dust close to the star, which we model as an 
unresolved 100\,K blackbody.  It would be difficult to reproduce this short-wavelength 
excess emission using only small dust grains, since the maximum temperature of 
astrosilicate grains at the radius of the scattered light ring is 70\,K, and furthermore
very small grains are ruled out by the lack of solid state features in the {\it Spitzer} 
IRS spectrum.  However, the resolved observations confirm that 
even the outer belt is significantly hotter than its blackbody equilibrium 
temperature.  As discussed by \citet{boo12}, this mismatch in sizes is typical 
of the effort to deduce debris disk sizes from SED fitting; in the 
particular case of HD 61005, increasing the porosity of the grains could help 
account for the high temperatures indicated by the SED.  It is also interesting 
to note that some polarimetric observations of debris disks seem to require 
high grain porosity to explain the observed properties of the scattered light 
\citep[e.g.,][]{gra07}.

\subsection{Comparison of Millimeter and Scattered Light Morphology}
\label{sec:structure}

When considering the morphology of the millimeter-wavelength emission, the
most salient question is how it compares with the structure observed in 
scattered light.  Based on the ISM-driven mechanism for creating the 
swept-back wings of the Moth \citep{man09,deb09} and the size-dependent
response of dust grains to such a mechanism \citep{mar11}, we expect
that the millimeter grains would be confined to the thin parent-body ring
at 61\,AU radius and absent from the swept-back features.  Indeed, the 
analysis in Section~\ref{sec:wings} demonstrates that a scenario in which 
the millimeter emission traces the morphology of the scattered light wings 
is ruled out at the 4\,$\sigma$ level.  While this analysis is rudimentary, it 
is certainly suggestive that we are observing wavelength-dependent 
structure.  The small grains responsible for scattered light appear to form 
the bulk of the material in the streamers, while the larger grains that dominate 
the 1.3\,mm emission are confined primarily to the narrow ring observed 
in scattered light.

Due to the relatively low signal-to-noise of the SMA observations we are as yet 
unable to decisively rule out an alternative scenario, in which millimeter 
grains at least to some extent trace the spectacular scattered light wings.  
If some of the emission from large grains does in fact originate from the 
wings, it would provide an indication that large grains are perturbed (to a 
lesser extent) by the same mechanism that is responsible for the swept-back 
structure observed in scattered light.  Given the theoretical work by 
\citet{mar11} indicating that large grains should remain unperturbed by ISM 
interactions, this would suggest that an alternative mechanism is responsible 
for the wings of the Moth.  An eccentric perturber -- for example, an unseen 
planet embedded in the disk -- may also be capable of breaking the symmetry 
or the disk and causing the bowl-shaped appearance of the scattered light 
(M. Fitzgerald, in prep).  Secular interactions can cause grains to acquire both
a forced eccentricity \citep{wya99} and inclination, the magnitude of which is
dependent on the ratio $\beta$ of radiation pressure to gravity.  Larger grains
should therefore exhibit less extreme swept-back structure, although the offset
of the disk center from the star position should persist at millimeter wavelengths.
Higher-resolution observations of the millimeter emission would therefore be
advantageous in order to further disambiguate the physical processes shaping 
the disk.  

%\subsection{Possible Role of Planets Shaping Disk Morphology}

%Streamer Grid

The somewhat ambiguous morphology of the millimeter-wavelength emission should
be considered in the context of hints that the HD 61005 system may 
host at least one planet.  The system has been considered from several 
different planet-related perspectives.  \citet{set08} included HD 61005 in 
an RV search for planetary companions around nearby young star, but report 
that the observed variability is consistent with stellar activity rather than 
planets; however, the length of the survey does not appear to be sufficient 
to detect planets orbiting at tens of AU from the central star.  \citet{wat11} 
include HD 61005 in a sample of debris disks with known viewing geometries, 
for which they estimate the inclination of the stellar rotation to the line 
of sight and search for evidence of misalignment.  They find that debris disks, 
including HD 61005, are generally well aligned with the rotation axes of 
their host stars.  The \citet{bue10} scattered light study indicates that 
the ring is offset from the star position by at least 2.75$\pm$0.85\,AU 
(in projection onto the sky plane, along the major axis only), indicative of eccentricity.  They also 
note a pronounced brightness asymmetry between the NE and SW components of 
the ring that is almost certainly due to density enhancements.  The upper 
limit on companion mass in the LOCI image is below the deuterium-burning 
limit, varying between roughly 3 and 6\,M$_\mathrm{Jup}$ between the inner 
working angle and the ring radius.  It is not yet clear whether the position 
offset and density enhancements could be caused by a planet, or whether they 
could be caused by the same ISM interaction that might be producing the 
streamers.  However, the 1.6\,$\mu$m wavelength of the VLT observations 
presented in \citet{bue10} approaches the grain population discussed in 
\citet{mar11}, which is too large to be effectively sculpted by the ISM.  
The near-IR eccentricity and asymmetric density distributions are suggestive
that a mechanism other than the ISM, possibly including planets, may be needed
to explain all the observed features of the system.  However, it should be 
emphasized that the millimeter emission alone is so far fully consistent with 
an ISM-sculpted disk morphology.

\section{Summary and Conclusions}
\label{sec:conclusions}

We have spatially resolved the dust continuum emission from the debris disk 
around HD 61005 at a wavelength of 1.3\,mm.  We observe a double-peaked 
structure consistent with an optically thin disk viewed close to edge-on.  
A simultaneous analysis of the spectral energy distribution and 
millimeter-wavelength visibilities demonstrates that the dust is hotter than the
expected blackbody equilibrium temperature given the relatively large radial 
extent of the ring resolved in scattered light and millimeter continuum emission.  
This is indicative of the presence of a substantial quantity of small grains, some 
of which are likely close to the blow-out size.  

We also investigate the morphology of the millimeter-wavelength emission,
particularly in comparison with the scattered light observations.  Our 
MCMC analysis suggests that the millimeter emission arises from roughly the
same stellocentric distance as the thin ring observed in scattered light.  
This is confirmed by a visibility-domain analysis of the millimeter emission
compared with a toy model based on the observed features of the scattered light
emission.  In the context of our toy model, a scenario in which the millimeter
morphology traces the scattered light flux distribution is ruled out at the
4$\sigma$ level.  This result is suggestive of wavelength-dependent structure, 
in which the large grains remain in the parent body ring, while small grains 
are preferentially affected by the perturbation responsible for sculpting the
scattered light wings.  Such grain size segregation is consistent with 
theoretical expectations for an ISM-sculpted disk.  The sensitivity and 
resolution of current observations is insufficient to provide a firm 
conclusion on the detailed morphology of the millimeter emission; hence these 
results remain merely suggestive.  However, these investigations pave the way 
for more sensitive future observations, for example with the Atacama Large 
Millimeter/Submillimeter Array (ALMA) currently nearing the end of its 
construction phase, and ripe to contribute to this exciting field.   

\acknowledgments We thank Holly Maness, who obtained the early data sets used 
in this paper and established detection.  We are also grateful to Esther 
Buenzli for kindly providing her LOCI image used in Figure~\ref{plot:Image}.  A.M.H. is 
supported by a fellowship from the Miller Institute for Basic Research in 
Science.

\bibliography{ms}

\end{document}